\documentstyle[12pt,epsfig,color]{article}
 \newcommand{\be}{\begin{eqnarray}}
 \newcommand{\ee}{\end{eqnarray}}
 \newcommand{\beq}{\begin{equation}}
 \newcommand{\eeq}{\end{equation}}
 \newcommand{\ba}{\begin{array}{1}}
 \newcommand{\ea}{\end{array}}
 \newcommand{\bb}{\begin{thebibliography}}
 \newcommand{\eb}{\end{thebibliography}}

 \newcommand{\re}[1]{(\ref{#1})}

 \newcommand{\abstitle}[1]{{\small {\bf #1}}}
 \newcommand{\absauthor}[1]{{\small {\bf #1}}}
 \newcommand{\address}[1]{{\it #1}}
 
\begin{document}
 \begin{center}
 \abstitle{Dielectron production in pion-nucleon reactions at intermediate
energies}\\
 \vspace{0.3cm}
 \absauthor{
A.P.Ierusalimov, G.I.Lykasov }\\ [2.0mm]
 \address{
JINR, Dubna, Moscow region, 141980, Russia }
\end{center}
 \vspace{0.2cm} 
\vskip 5mm
\begin{abstract}
  \textcolor{black}{Dielectron} production in the $\pi N$ interaction at not large energies is
  studied. The dominant contribution of the $\Delta$-isobar creation in the intermediate state at
  incident pion momenta \textcolor{black}{of} about 0.3-0.4 GeV$/$c is shown.
  The \textcolor{black}{experimental distributions} over the angle and effective mass $M_{e^+e_-}$ of 
  \textcolor{black}{the} $e^+e^-$ pair are described satisfactorily. \textcolor{black}{This}
  stimulated us to present \textcolor{black}{theoretical} predictions for the $M_{e^+e_-}$
  distribution in the process $\pi^- p\rightarrow ne^+e^-$ at different incident momenta, which
  could be verified, for example, by \textcolor{black}{the} HADES experiments.      
\end{abstract}

 
\vspace{1cm}\vspace{1cm}
\section{Introduction}
Processes of meson electroproduction play an important role in
\textcolor{black}{the} study of the structure and properties of matter,
see, for example, the review \textcolor{black}{in ref.} \cite{Amaldi:1979} and
references therein.
The production of dileptons in hadron-hadron, hadron-nucleus and
nucleus-nucleus reactions \textcolor{black}{has been} analyzed very intensively
\cite{Blokhintseva:1999}-\cite{Kaptari:2006}. In these reactions, virtual
photons, which materialize \textcolor{black}{as} $e^+e^-$ \textcolor{black}{pairs}, 
carry unique information on \textcolor{black}{the} properties of matter and the
reaction mechanism.

\textcolor{black}{Inverse} pion electroproduction (IPE), $\pi N\rightarrow e^+e^- N$,
can \textcolor{black}{provide} information on the nucleon electromagnetic structure
in the time-like region. \textcolor{black}{An} intense pion beam could allow to perform
more detailed experiments on the IPE aimed at both extracting the hadron structure
and carrying out a multipole analysis \textcolor{black}{like in the case of}
photoproduction and electrproduction, see for example \cite{Kamalov} and
references therein.     
       
\section{General formalism}

\begin{figure}[ht]
{\epsfig{file= 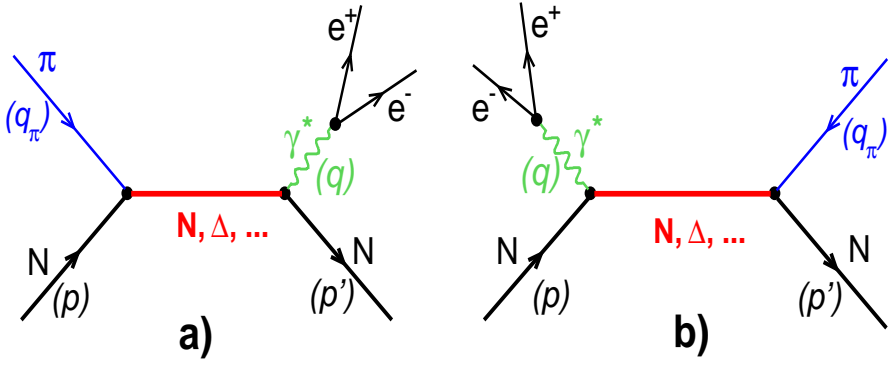,height=5.cm,width=12.cm  }}
\caption[Fig.1]{The one-nucleon or one-nucleon resonance exchange graph in the $s$-channel of
the IPE $\pi N\rightarrow\gamma N\rightarrow e^+e^- N$ process.}
\label{Fig.1}
\end{figure}

\begin{figure}[ht]
{\epsfig{file= 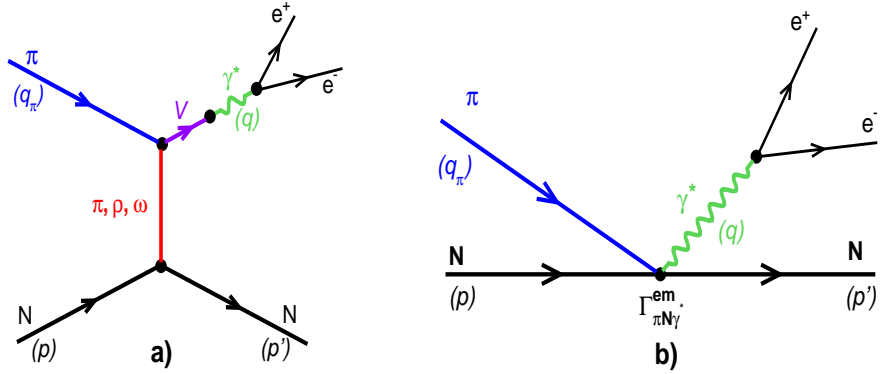,height=5.cm,width=12.cm  }}
\caption[Fig.2]{The one-meson exchange graph in the $t$-channel (a)
and the electromagnetic contact term (b) for the IPE.}
\label{Fig.2}
\end{figure}

We analyze the reaction $\pi N\rightarrow\gamma^* N\rightarrow e^+e^- N$ within the
unified model. \textcolor{black}{This} means that in the one-photon approximation,
\textcolor{black}{owing to} $T$-invariance, three reactions $\gamma N\rightarrow\pi N,
e N\rightarrow e\pi N$ and $\pi N\rightarrow e^+e^- N$ are related to the process 
$\gamma^* N\leftrightarrow\pi N$ \textcolor{black}{by} the hadron current
$J_\mu(s,t,m_\gamma^2)$, where $m_\gamma^2=0,>0$ and $m_\gamma^2<0$ correspond 
\textcolor{black}{to} pion photoproduction, electroproduction and inverse pion
electroproduction (IPE), respectively \cite{Berends:1967,Surovtsev:2005}.
For the IPE the $S$-matrix in the one-photon approximation reads \textcolor{black}{as}
\begin{equation}
S_{fi}=\delta_{fi}+(2\pi)^4i\delta^{(4)}(p+q\pi-q-p^\prime)\frac{m}{\sqrt{2q_{\pi 0}p_0q_0p^\prime_0}}
\epsilon^\mu J_\mu(s,t,m_{\gamma^*}^2)~,
\label{def:Sfi}
\end{equation}
where $\epsilon^\mu$ is the electromagnetic current and photon propagator
\begin{equation}
\epsilon^\mu~=~{\bar u}(k_2)\gamma^\mu u(k_1)e/q^2
\label{dev:epsilmu}
\end{equation}  
According to \cite{Berends:1967}, the current $J_\mu$ can \textcolor{black}{also be}
presented in the following \textcolor{black}{form:}
\begin{equation}
J_\mu~=~\sum_{i=1}^8 B_i(s,t,u,q^2)N_\mu^i~,
\label{def:Jmu}
\end{equation}
where the functions $N_\mu^i$ related to the Dirac $\gamma_\mu$ matrices are presented
in the Appendix, see also \cite{Berends:1967}. Inserting functions $N_\mu^i$ into
Eq.~\ref{def:Jmu} one can \textcolor{black}{obtain} the following form for the current
$J_\mu$: 
\begin{equation}
J_\mu~=~\gamma_5[G_1P_\mu+G_2q_{\pi\mu}+G_3q_\mu+G_4\gamma_\mu+G_5{\hat q}P_\mu+G_6{\hat q}q_{\pi\mu}
+G_7{\hat q}q_\mu+G_8{\hat q}\gamma_\mu]~,
\label{def:JmuG}
\end{equation} 
where ${\hat q}={\tilde q}^\nu\gamma_\nu$, ${\tilde q}=q/|q|$ is the
\textcolor{black}{unit photon} four-momentum~;
\be
B_1=-iG_8~,~B_2=2iG_1~,~B_3=2iG_2~,~B_4=2i(G_3+2G_8)~,\\
\nonumber
~B_5=G_4~,~B_6=G_5~,~B_7=G_7~,~B_8=G_6~.
\label{def:relBG}
\ee
Using the current conservation \textcolor{black}{law}
\begin{equation}
q^\mu J_\mu=0~,
\label{def:consJmu}
\end{equation}
we \textcolor{black}{actually obtain} 6 independent functions instead of 8,
see \cite{Berends:1967}. The forms for functions $G_i$ depending on the Mandelstam
relativistic invariants $s,t$ and $u$ are obtained calculating the Feynman graphs
presented in Figs.~(\ref{Fig.1},\ref{Fig.2}).

On the other \textcolor{black}{hand}, \textcolor{black}{applying} the current
conservation \textcolor{black}{law} given by Eq.(\ref{def:consJmu}) one can
\textcolor{black}{obtain} the following form for $\epsilon^\mu J_\mu$, as a sum of
six independent functions $A_i(s,t,u,q^2)$:
\begin{equation}
\epsilon^\mu J_\mu~=~\sum_{i=1}^6 A_i(s,t,u,q^2)M_\mu^i~
\label{dev:epsJmu}
\end{equation} 
where the functions $M_\mu^i$ are presented in \cite{Berends:1967} and $A_i(s,t,u,q^2)$
are related to the functions $B_i(s,t,u,q^2)$ entering into
Eq.(\ref{dev:epsJmu})\textcolor{black}{,}
\be
A_1=B_1-mB_6~,~A_2=\frac{2B_2}{t-\mu^2}~,~A_3=-B_8~,~A_4=- \frac{1}{2}B_6~,\\
\nonumber
A_5=\frac{1}{q\cdot q_\pi}(B_1+2B_4-\frac{2P\cdot q}{t-\mu^2})~,~A_6=B_7~,
\label{def:AiBi}
\ee

The current $J_\mu$ can \textcolor{black}{also be presented in terms of} Pauli matrices
\cite{Blokhintseva:1999,Berends:1967,Tiator:1992}, see Appendix.
\section{One-nucleon exchange graph}
$\bullet~${\bf $s$-channel one-nucleon exchange graph}\\
Calculating the \textcolor{black}{one-nucleon exchange} Feynman graph in the $s$-channel (Fig.1a), the so-called 
Born-term,
we use the effective meson-NN Lagrangian in the form (see \cite{Ericson:1988,Shyam:2003})
\begin{equation}
{\cal L}_{NN\pi}~=~-\frac{g_{NN\pi}}{2m_N}{\bar\Psi}_N\gamma_5\gamma_\mu{\bf\tau}\cdot
(\partial^\mu\Phi_\pi)\Psi_N~,
\label{def:piNNLagr}
\end{equation}
where $\Psi_N$ is the spinor for a nucleon and $\Phi_\pi$ is the pion field.
We use the monopole form factor in this interaction vertex
\begin{equation}
F_\pi^{NN}=\frac{\Lambda_\pi^2}{\Lambda_\pi^2+q^2}~,
\end{equation} 
where $\Lambda$ is the cut-off parameter. The usual nucleon propagator is used
\begin{equation}
S^{(s)}_N(p_N)=\frac{\not{p}_N+m_N}{p_N^2-m_N^2}=\frac{\not{p}^\prime+\not{q}+m_N}{s-m_N^2}
\equiv\frac{\not{p}+\not{q}_\pi+m_N}{s-m_N^2}~,
\end{equation}
where $\not{p_N}=p_N^\mu\cdot\gamma_\mu$, $p_N=p^\prime+q=q_\pi+p$ is the four-momentum of the 
exchanged 
nucleon, $p_N^2=s=(p+p_\pi)^2=(p^\prime+q)^2$ $m_N$ is the nucleon mass, $\gamma_\mu$ is the Dirac 
matrix.\\
The general form for the effective Lagrangian for the nucleonic \textcolor{black}{bremsstrahlung}
process $NN\gamma$ reads \textcolor{black}{as} 
\begin{equation}
{\cal L}_{NN\gamma}~=~-e{\bar\Psi}_N\Gamma^{NN\gamma}_\mu A^\mu \Psi_N~,
\end{equation}
where $A^\mu$ is the electromagnetic field, the half-off-shell nucleon-photon vertex
$\Gamma^{NN\gamma}_\mu$ is \cite{Shyam:2003}
\begin{equation}
\Gamma^{NN\gamma}_\mu~=~-ie\sum_{s=\pm}(F_1\gamma_\mu+F_2\Sigma_\mu+F_3q_\mu)\Lambda_s~,
\end{equation} 
where $\Lambda_\pm=(\pm{\not{p}}+W)/W$ are the projection operators for the off-shell
nucleon \textcolor{black}{of} mass $W=\sqrt{p^2}$ and $\Sigma_\mu=i\sigma_{\mu\nu}q^\nu/(2m)$.
Here we use the on-shell form factors for the $NN\gamma$ vertex and assume that
$F_1^+=F_1^-=F_1, F_2^+=F_2^-=F_2$ and $F_3^+=F_3^-=0$, as it was suggested in
\cite{Shyam:2003,Mosel:1994,Haglin:1989}. This assumption can be rather applicable at
not large values of the initial pion. We will \textcolor{black}{demonstrate this for a}  
pion momentum \textcolor{black}{of} about 0.3 GeV/c  
\\
$\bullet~${\bf $u$-channel one-nucleon exchange graph}\\
For the one nucleon-exchange graph in the $u$-channel (Fig.1b) the propagator reads
\begin{equation}
S^{(u)}_N(p_N)=\frac{\not{p}^\prime-\not{q}+m_N}{u-m_N^2}
\equiv\frac{\not{p}-\not{q}_\pi+m_N}{u-m_N^2}~,
\end{equation}
where $u=(q_\pi-p^\prime)^2=(p-q)^2$. We should \textcolor{black}{also substitute} the
four-momenta $q$ and $q_\pi$ \textcolor{black}{by} $-q$ and $-q_\pi$ in
\textcolor{black}{all the} vertices of this graph. 
\section{$\Delta$-isobar exchange graph}
\textcolor{black}{In calculating} the $\Delta$-isobar exchange graphs in the $s$-
and $u$-channels we use the same procedure \textcolor{black}{as} in
\cite{Shumeiko:1973,Shyam:2003,Kaptari:2006}. In particular, the $\Delta$-nucleon-photon
vertex is
\begin{equation}
\Gamma^{\mu\nu}_{\Delta N\gamma}=-i\frac{e}{2m}\left\{g_1(q^2)\gamma_\lambda+
\frac{g_2(q^2)}{2m}p^\Delta_\lambda+\frac{g_3(q^2)}{2m} q_\lambda\right\}
(-q^\nu g^{\mu\lambda}+q^\mu g^{\nu\lambda})~,
\end{equation} 
where $g_1,g_2,g_3$ are the form factors used in the same monopole form and the coupling
constants \textcolor{black}{are} like in \cite{Shyam:2003}.
We use the following effective $\Delta-nucleon-meson$ Lagrangian \cite{Shyam:2003,Feuster}:
\begin{equation}
{\cal L}_{\Delta N\pi}=\frac{g_{\Delta N\pi}}{mu_\pi}
\Psi_\mu^\Delta\gamma_5\partial^\mu\Phi_\pi\Psi_N+H.c.~,
\end{equation}
where $\Psi_\mu^\Delta$ is the vector spinor for the $\Delta$-isobar.
The propagator for the $\Delta$-isobar reads 
\be
S_\Delta^{\mu\nu}(p_\Delta)=-\frac{i(\not{p}_\Delta+m_\Delta)}{p_\Delta^2-
(m_\Delta-i(\Gamma_\Delta)/2)^2}(g^{\mu\nu}-\frac{1}{3}\gamma^\mu\gamma^\nu-
\frac{2}{3m_\Delta^2}p_\Delta^\mu p_\Delta^\nu+ \\
\nonumber
\frac{1}{3m_\Delta}(p_\Delta^\mu\gamma^\nu
-p_\Delta^\nu\gamma^\mu))
\ee 
where $p_\Delta$ and $M_\Delta$ are the four-momentum and the mass of the
$\Delta$ isobar created in the intermediate state (Figs.1(a,b)) and the partial 
width of the off-shell $\Delta$-isobar $\Gamma_\Delta\equiv\Gamma_{\Delta\rightarrow\pi N}$ 
is \cite{Shyam:2003,Wolf:1990,Teis:1997,Kagarlis:2000}
\begin{equation}
\Gamma_\Delta(M)=\Gamma_0\left(\frac{k^*(M)}{k^*(M_\Delta)}\right)^3
\frac{M_\Delta}{M}
\left(\frac{\beta^2+(k^*(M_\Delta))^2}{\beta^2+(k^*(M))^2}\right)^2~.  
\label{def:deltapart}
\end{equation}
Here $\Gamma_0=0.12 GeV$ is the bar width of $\Delta$, $\beta$ is the parameter, we \textcolor{black}{assumed}  
$\beta^2=0.25 (GeV/c)^2$ as in \cite{Shyam:2003}, $k^*$ is the pion momentum in the $\pi-N$
c.m.s. and
$k^*(M)=\lambda^{1/2}(M^2,\mu_\pi^2,m^2)/2M$
\section{One-meson exchange in $t$-channel}
Calculating the \textcolor{black}{Feynman} graphs corresponding to the one-meson exchange in the 
$t$-channel, Fig.2a, we \textcolor{black}{only included the} one-pion exchange
\textcolor{black}{so as to} analyze the IPE processes at intermediate energies
less than 0.5-0.6 GeV when the vector meson exchange graphs
\textcolor{black}{contribute insignificantly} and can be neglected.
\textcolor{black}{Moreover,} 
note that the $G$-parity arguments forbid $\omega\pi\pi$ and
$A_1\pi\pi$ vertices, so the pion current diagram (Fig.2a) only contributes 
\textcolor{black}{in the case of} isovector currents.
The $NN\pi$ Lagrangian was taken in the form given by eq.(\re{def:piNNLagr}),
whereas the Lagrangian of the $\rho\pi\pi$ interaction was taken in the following
form \cite{Towner:1987}:     
\begin{equation}
{\cal L}_{\rho\pi\pi}=g_\rho\epsilon_{jlm}\Phi_{\rho,\mu}^j\Phi_\pi^l
\partial_\mu\Phi_\pi^m~,
\label{def:rhopipi}
\end{equation}
where $\Phi_\rho$ and $\Phi_\pi$ are the $\rho$-meson and pion fields respectively,
$\epsilon_{jlm}$ is the fully antisymmetric unit tensor. The coupling constant 
$g_\rho\equiv g_{\rho\pi\pi}$ was found from the $SU(3)$ relation \cite{Towner:1987}
$2g_{\rho NN}=g_\rho$, whereas the coupling constant $g_{\rho NN}$ was taken from
\cite{Machleidt:1987}. 
The Lagrangian corresponding to the $\rho\gamma$ vertex (Fig.2a) was taken in the
following form \cite{Kaptari:2006}: 
\begin{equation}
{\cal L}_{\rho\gamma}^{em}=-\frac{e}{2f_{\rho\gamma}}{\cal F}^{\mu\nu}
{\cal G}_{\nu\mu}^\rho~,
\label{def:rhogamma}
\end{equation}
where ${\cal F}^{\mu\nu}$ and ${\cal G}^{\nu\mu}$ are the strength tensors 
for the photon and $\rho$-meson fields.
For the $\rho$-meson propagator $S_\rho^{\mu\nu}$ of Fig.2a we \textcolor{black}{obtained}
the same form as in \cite{Shyam:2003}\textcolor{black}{,}
\begin{equation}
S_\rho^{\mu\nu}(q^2)=-i\left(\frac{g^{\mu\nu}-q^\mu q^\nu /q^2}{q^2-m_\rho^2}
\right)~,
\label{def:Srho}
\end{equation}
where $m_\rho$ is the $\rho$-meson mass.
The pion propagator (Fig.2a) reads
\begin{equation}
S_\pi(t)=\frac{i}{t-\mu_\pi^2}~,
\label{def:Spi}
\end{equation}
where $t=(q_\pi-q)^2$.
\section{Electromagnetic contact term}
When the pseudevector $\pi NN$ interaction given by Eq.(\ref{def:piNNLagr}) is used
then \textcolor{black}{the} contact electromagnetic diagram (Fig.2b) should be included. 
The electromagnetic contact terms, e.g., the vertices with four lines as in
Fig.2b\textcolor{black}{,} \cite{Kroll-Ruder,Kaptari:2006,Marcucci:1998} correspond to the
following interaction Lagrangian:
\begin{equation}
{\cal L}_{NN\pi\gamma}~=~-\frac{{\hat e}f_{NN\pi}}{\mu_\pi}
{\bar\Psi}_N\gamma_5\gamma_\mu A_\mu{\bf(\tau}\cdot
\Phi_\pi)\Psi_N~,
\label{def:contterm}
\end{equation}
where ${\hat e}$ is the charge operator of the pion.
\section{Amplitude and cross section}
In the one-photon approximation the dielactron production in the IPE process is considered
as a decay of the virtual photon produced in strong and electromagnetic $\pi-N$ interactions.
These processes and the similar $NN\rightarrow\gamma^* NN\rightarrow e^+e^- NN$ reactions
\textcolor{black}{have already been} investigated, see for
example \cite{Berends:1967}-\cite{Kaptari:2006}, where all the details to obtain the forms
for the amplitude and differential cross sections are presented. The general expression
for the IPE relativistic invariant amplitude squared reads
\begin{equation}
\mid{\cal T}\mid^2~=~W_{\mu\nu}\frac{e^2}{q^2}l^{\nu\mu}~,
\end{equation}  
where $e$ is the electron charge, $q^2=m_{\gamma^*}^2=(k_1+k_2)^2$, $q$ is the four-Momentum of the
virtual time-like photon decaying into $e^+$ and $e^- $ with the four-Momenta $K_1$ and $k_2$ 
respectively;
$W_{\mu\nu}=\sum_{spins}J_\mu J_\nu^+$ is the hadronic tensor, whereas $l^{\nu\mu}$ is the leptonic 
tensor.
\begin{equation}
l_{\nu\mu}=4(k_{1\nu}k_{2\mu}+k_{1\mu}k_{2\nu}-g_{\nu\mu}(k_1\cdot k_2))
\end{equation}
where $k_1$ and $k_2$ are the four-momenta of the outgoing electron and
positron\textcolor{black}{,} respectively. 
\textcolor{black}{Following the procedure} to get the differential cross
section presented in \cite{Kaptari:2006} one can \textcolor{black}{obtain} the following forms for the
effective mass distribution $d\sigma/dM_{e^+e^-}$ of the $e^+ e^-$ pair and the
\textcolor{black}{angular} distribution $d\sigma/d\cos\theta_{\gamma^*}^*$ 
of the virtual $\gamma^*$ in the $\pi-N$ c.m.s.:
\begin{equation}
\frac{d\sigma}{d M_{e^+e^-}}~=~\frac{\alpha_{em}^2}{12\pi M_{e^+e^-}s\lambda^{1/2}(s,\mu_\pi^,m^2)}
\int_{t^-}^{t^+}\frac{\lambda^{1/2}(s,q^2,m^2)}{s}\sum_{spins}J_\mu J_\nu^+ dt~;
\label{def:effmassdist}
 \end{equation} 
 \begin{equation}
\frac{d\sigma}{d \cos\theta_{\gamma^*}^*}~=~\frac{\alpha_{em}^2}{48\pi s\lambda^{1/2}(s,\mu_\pi^,m^2)}
\int\frac{\lambda^{1/2}(s,q^2,m^2)}{q^2}\sum_{spins}J_\mu J_\nu^+dq^2~,
\label{def:costetadist}
 \end{equation} 
where $\alpha_{em}=e^2/4\pi=1/137$; $ \lambda(x^2,y^2,z^2)=(x^2-(y+z)^2)=(x^2-(y-z)^2)$.
\section{Dalitz decay of $\Delta$ resonance}
The effective mass \textcolor{black}{distribution} of the $e^+e^-$ pair produced in the
$\pi-N$ reaction at \textcolor{black}{an} energy corresponding to the $\Delta$ isobar
creation in the intermediate state can be estimated applying the simple Breit-Wigner
approximation and the Dalitz decay of \textcolor{black}{the} $\Delta$ isobar\textcolor{black}{,}  
\cite{Kagarlis:2000} 
\begin{equation}
\frac{d\sigma_{\pi N}^{Br.W.}}{dM_{e^+e^-}}=\frac{1}{3}\frac{4\pi}{(k^*)^2}\frac{M_\Delta^2
\Gamma_{\Delta}}
{(M_\Delta^2-s-iM_\Delta\Gamma_{tot})^2}\frac{d\Gamma_{\Delta\rightarrow\gamma^* N
\rightarrow e^+e^- N}}
{dM_{e^+e^-}}~,
\label{def:BR-Wign}
\end{equation}
where $\Gamma_{\Delta}$ is given by Eq.(\ref{def:deltapart}) and
\begin{equation}
 \frac{d\Gamma_{\Delta\rightarrow\gamma^* N\rightarrow e^+e^- N}}{dM_{e^+e^-}}=
\frac{2\alpha_{em}}{3\pi M_{e^+e^-}}\sqrt{1-\frac{4\mu_e^2}{M_{e^+e^-}^2}}\left(1+
\frac{2\mu_e^2}{M_{e^+e^-}^2}\right)\Gamma_{\Delta\rightarrow\gamma^* N}~;
\end{equation}
\begin{equation}
\Gamma_{\Delta\rightarrow\gamma^* N}=\frac{p^*_\Delta}{8\pi m_\Delta^2}
\mid{\cal M}_{\Delta\rightarrow\gamma^* N}\mid^2~.
\end{equation}
Here $p^*_\Delta$ is the momentum in the $\gamma^* N$ c.m.s. and the matrix element squared reads
\be
\mid{\cal M}_{\Delta\rightarrow\gamma^* N}\mid^2=e^2G_M^2\frac{(m_\Delta+m)^2((m_\Delta-m)^2-
M_{e^+e^-}^2)}{4m^2((m_\Delta+m)^2-M_{e^+e^-}^2)}
(7m_\Delta^4+14m_\Delta^2M_{e^+e^-}^2+\\
\nonumber
3M_{e^+e^-}^4+8m_\Delta^3 m+2m_\Delta^2m^2+6M_{e^+e^-}^2m^2+3m^4)~.
\ee
The \textcolor{black}{angular} distribution of $\gamma^*$ is presented in \cite{Kagarlis:2000}. 
\section{Results and discussion}
\begin{figure}[ht]
{\epsfig{file=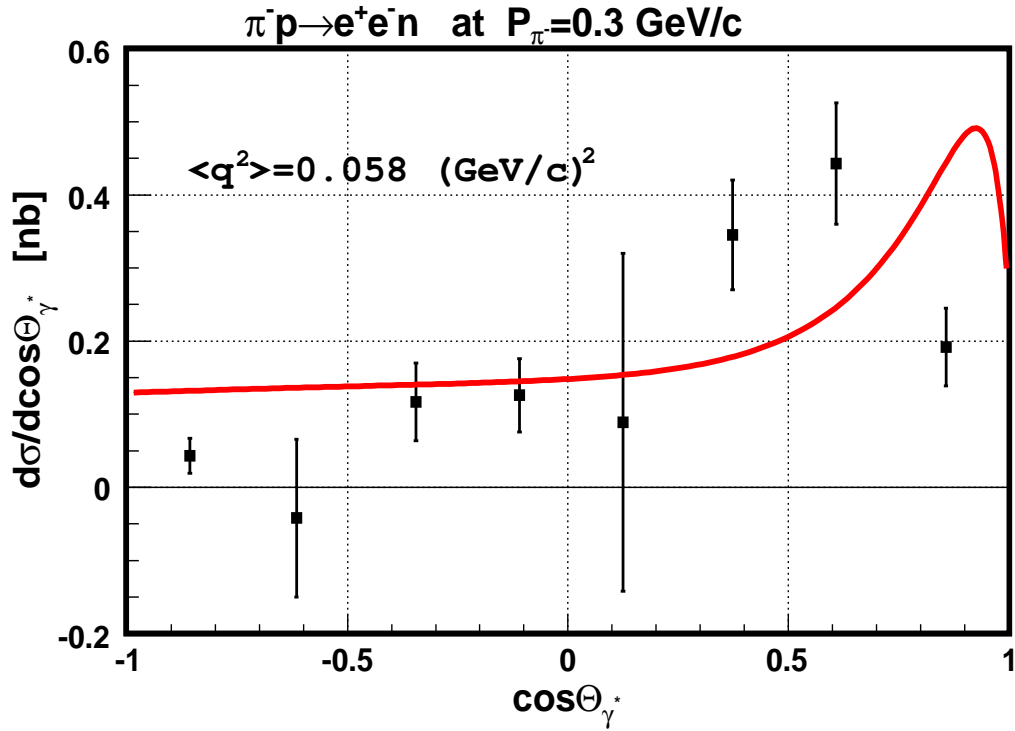,height=10.cm,width=14.cm  }}
\caption{\textcolor{black}{Angular} distribution $d\sigma/d\cos\theta_{\gamma^*}$, 
where $\theta_{\gamma^*}$ is the angle of the virtual photon $\gamma^*$. 
The solid curve corresponds to our calculations. The experimental
data are taken from \cite{Berezhnev:1976}.}
\label{Fig.3}
\end{figure}

We \textcolor{black}{have} calculated the distributions of $e^+e^-$ \textcolor{black}{pairs} as
\textcolor{black}{functions} of the dielectron effective mass $M_{e^+e^-}$ and
\textcolor{black}{of} $cos\theta_{\gamma^*}$\textcolor{black}{,} respectively, see
Eqs.~(\ref{def:effmassdist},\ref{def:costetadist}), adding incoherently the Dalitz decay
of the $\Delta$ resonance. 
In Fig.~\ref{Fig.3} the \textcolor{black}{angular} distribution $d\sigma/d\cos\theta_{\gamma^*}$,
where $\theta_{\gamma^*}$ is the angle of the virtual photon $\gamma^*$,
is presented. This distribution was summed over $q^2$  in the interval
$0.046(GeV/c)^2\ge q^2\le 0.065(GeV/c)^2$, according to the experimental 
data \cite{Berezhnev:1976}. The solid curve corresponds to our calculations, e.g., the
coherent sum of the one-nucleon and $\Delta$-exchange graphs with the positive phase sign
of the second graph. 
The experimental data are taken from \cite{Berezhnev:1976}.
It is seen from this figure that our total calculation  
results in \textcolor{black}{a} more or less isotropic distribution and provides the qualitative 
description of the data, which have big error bars.

\begin{figure}[ht]
{\epsfig{file=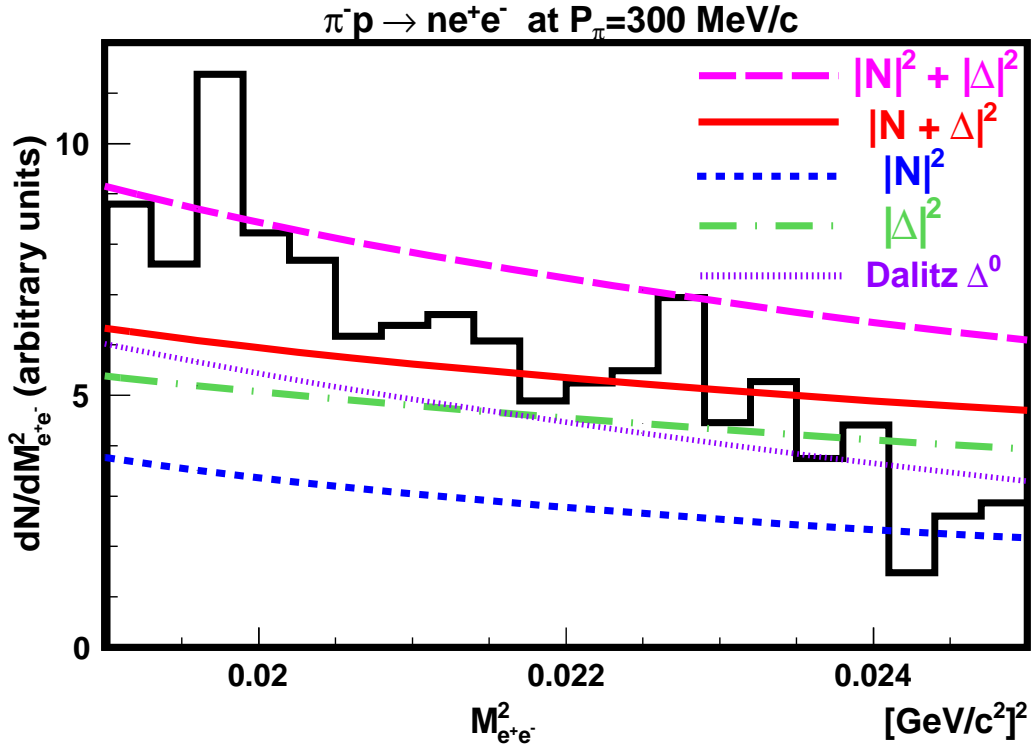,height=10.cm,width=14.cm  }}
\caption{\textcolor{black}{Invariant} mass distribution for 
dielectrons\textcolor{black}{,} $d\sigma/dM^2_{e^+e^-}$\textcolor{black}{,} 
produced in the reaction $\pi^- p\rightarrow e^+e^- n$. 
The histogram \textcolor{black}{presents} the experimental data \textcolor{black}{in arbitrary}
units \cite{Hoffman:1983}. 
Contributions of the Dalitz decay, the one-nucleon exchange and the $\Delta$-isobar exchange 
correspond to the dotted, short dashed and dash-dotted curves respectively.    
The solid curve is our calculation of the coherent sum of all diagrams of Figs.~(\ref{Fig.1},\ref{Fig.2}),
the dashed line corresponds to the incoherent sum of the one-nucleon and $\Delta$-exchange contributions.} 
\label{Fig.4}
\end{figure}

In Fig.~\ref{Fig.4} the invariant mass distribution for dielectrons $d\sigma/dM^2_{e^+e^-}$
produced in the reaction $\pi^- p\rightarrow e^+e^- n$ is presented. The histogram is 
the experimental data in arbitrary units \cite{Hoffman:1983}.
The solid curve is our total calculation , e.g. the coherent sum of all the graphs of
Figs.(1,2). The dash-dotted curve is the contribution of $\Delta$-exchange diagrams of
Fig.1, while the dotted curve corresponds to the calculation using the Breit-Wigner
form for the spectrum given by Eq.(\ref{def:BR-Wign}) including the Dalitz decay of
$\Delta$. The long dashed line corresponds to the incoherent sum of the one-nucleon and 
$\Delta$ isobar exchange diagrams. \textcolor{black}{The} Dalitz
decay of resonances was calculated, according to \cite{Kagarlis:2000}, see also
\cite{BrCass:08}. 

\begin{figure}[ht]
{\epsfig{file=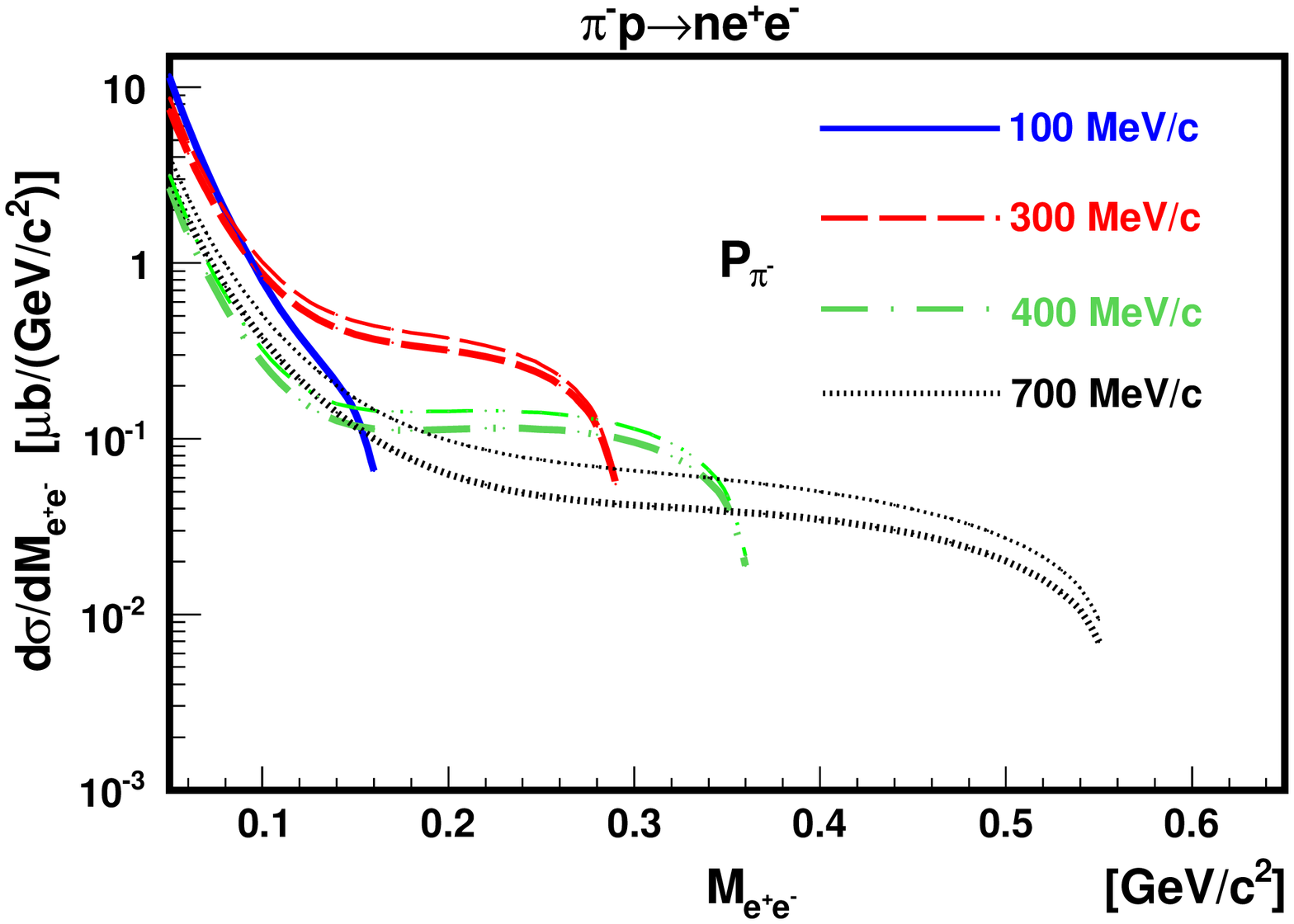,height=10.cm,width=14.cm  }}
\caption{\textcolor{black}{Invariant} mass distribution for  
dielectrons\textcolor{black}{,} $d\sigma/dM_{e^+e^-}$\textcolor{black}{,} 
produced \textcolor{black}{in reaction} $\pi^- p\rightarrow e^+e^- n$ at
different pion momenta. 
The thick curves correspond \textcolor{black}{to calculations} without any form
factors $F(q^2)$ or $F(t)$, while the thin lines correspond
\textcolor{black}{to inclusion} of the form factors in the vertices of diagrams
\textcolor{black}{in} Figs.~(\ref{Fig.1},\ref{Fig.2}).}    
\label{Fig.5}
\end{figure}

In Fig.~\ref{Fig.5} the invariant mass distribution for  
dielectrons\textcolor{black}{,} $d\sigma/dM_{e^+e^-}$\textcolor{black}{,} 
produced \textcolor{black}{in reaction} $\pi^- p\rightarrow e^+e^- n$ at
different pion momenta. 
The solid, dashed, dash-dotted and dotted curves correspond \textcolor{black}{to calculations} 
at the initial pion momentum about 100 MeV$/$c, 300 MeV$/$c, 400 MeV$/$c and 700 MeV$/$c
respectively.
The thick lines correspond to our calculations without any form  
factors $F(q^2)$ (or $F(t)$), while the thin curves correspond
\textcolor{black}{to inclusion} of the form factors in the vertices of diagrams
\textcolor{black}{in} Figs.~(\ref{Fig.1},\ref{Fig.2}).           

Let us note that our calculations were performed using the Feynmann diagrams of 
Figs.~(\ref{Fig.1},\ref{Fig.2}), which result in the main contribution to the observables
presented in Figs.~(\ref{Fig.4},\ref{Fig.5}), see for example, \cite{Gribov:1967}.
In our calculations we neglected the contributions of processes 
$\pi^- p \rightarrow \pi^0 p \rightarrow e^+e^-\gamma p$, $\pi^- p \rightarrow \eta^0 p$
and $\pi^- p \rightarrow \rho^0 p$ with subsequent decays ($\eta^0 \rightarrow e^+e^- 
\gamma$ and $\rho^0 \rightarrow e^+e^-$  respectively) because they can be significant
at initial momenta above 600-700 MeV$/$c and high dielectron effective masses 
$M_{e^+e^-}\geq$ 0.5 GeV$/$c$^2$, as it has been shown in \cite{pipiN,JL_np:2017}.   
\section{Conclusion}
In this paper we \textcolor{black}{have analyzed inverse} pion
electroproduction (IPE) processes $\pi N\rightarrow e^+e^- N$
\textcolor{black}{at intermediate} energies within
the Feynman graph formalism. The \textcolor{black}{dominant} contribution of the 
$\Delta$-isobar exchange graph in the $s$-channel to the effective 
mass spectrum of $e^+e^-$ \textcolor{black}{pairs} produced \textcolor{black}{at  
incident pion momenta of} about 0.3-0.4 GeV/c is shown, \textcolor{black}{while}
at higher momenta it decreases. At higher momenta the contributions
\textcolor{black}{of diagrams} with another baryon-exchange in the 
$s$-channel and \textcolor{black}{with a} vector meson-exchange in the
$t$-channel are \textcolor{black}{considerable}.  
We show \textcolor{black}{that careful} calculations of the Feynman graphs
\textcolor{black}{with account of} their interference \textcolor{black}{provides  
  correct} results for the observables which \textcolor{black}{may differ from
  the results of approximate calculations involving only bremsstrahlung}
of the nucleon and the Dalitz decay of baryon resonances in the intermediate
state. The existing experimental information on these processes is very poor. 
Therefore we \textcolor{black}{present predictions} for the $M_{e^+e^-}$-spectrum
and the \textcolor{black}{angular} distribution of the virtual photon decaying
into $e^+e^-$ in the IPE reactions \textcolor{black}{at incident momenta} less than
1.GeV/c. \textcolor{black}{This} can be verified \textcolor{black}{by future}
HADES experiments \textcolor{black}{with pion beams}. 
We \textcolor{black}{will  
extend} this approach to \textcolor{black}{analyze inelastic}
$\pi p\rightarrow e^+e^- X$ reactions \textcolor{black}{at HADES} energies.

{\bf  Acknowledgement\textcolor{black}{s}.}
We are grateful to T.D.Blokhintseva, T.Galatyuk, B.Friman, G.Pontecorvo, 
R.Holzman, S.Kamalov, V.P. Ladygin, U.Mosel, B.Ramstein, A.Rustamov, 
P.Salabura, J.Stroth, V.Shklyar, Yu.Surovtsev for very useful discussions.

\section{Appendix}
\be
N_1^\mu=i\gamma_5\gamma^\mu\hat{q}~;~N_2^\mu=2i\gamma_5 P^\mu~;~N_3^\mu=2i\gamma_5 q_\pi^\mu~;~
N_4^\mu=2i\gamma_5 q^\mu~;\\
\nonumber
N_5^\mu=\gamma_5\gamma^\mu~;~N_6^\mu=\gamma_5\hat{q} P^\mu~;~N_7^\mu=\gamma_5\hat{q} q^\mu~;
N_7^\mu=\gamma_5\hat{q} q_\pi^\mu
\ee
\be
{\bf J}=\frac{4\pi\sqrt{s}}{m}[i\tilde{\sigma} F_1+\sigma\cdot{\hat{\bf q}}_\pi(\sigma\times{\hat{\bf q}})
F_2+
i{\tilde{\bf q}}_\pi(\sigma\cdot\hat{\bf q})F_3+i{\hat{\bf q}}_\pi(\sigma\cdot{\hat{\bf q}}_\pi)F_4+
i{\hat{\bf q}}(\sigma\cdot{\hat{\bf q}})F_6]~,
\ee
where $\tilde{\sigma}=\sigma-{\hat{\bf q}}(\sigma\cdot{\hat{\bf q}_\pi})$ and
${\tilde{\bf q}_\pi}={\hat{\bf q}}_\pi-\hat{\bf q}({\hat{\bf q}}_\pi\cdot\hat{\bf q})$.
\textcolor{black}{The} $J^0$ component of the current $J^\mu$ can be
\textcolor{black}{obtained} from the current conservation \textcolor{black}{law}
given by Eq.(\ref{def:consJmu})
\be
J^0=\frac{4\pi\sqrt{s}}{m}i[(\sigma\cdot{\hat{\bf q}}_\pi)F_7+(\sigma\cdot{\hat{\bf q}})F_8]=
\frac{{\bf q}\cdot{\bf J}}{q^0}~.
\ee
\textcolor{black}{Functions} $F_7,F_8$ are related to $F_5,F_6$ \cite{Berends:1967,Tiator:1992}.
\be
\mid{\bf q}\mid F_5=q^0F_8~;~\mid{\bf q}\mid F_6=q^0F_7~
\ee
and functions $F_i, i=1-6$ can be obtained calculating the Feynmann graphs of
Figs.~(\ref{Fig.1},\ref{Fig.2}). 

\end{document}